\documentclass[sigconf]{acmart}
\usepackage{diagbox}
\usepackage{multirow}

\newcommand{\dR}{\nabla_{\tilde{\mathbf{R}}}}

\newcommand{\dw}{\nabla_{\mathbf{W_p}}}
\newcommand{\dz}{\nabla_{\mathbf{s}}}
\newcommand{\dzz}{\nabla_{\mathbf{\hat{s}}}}
\newcommand{\dfz}{\nabla_{\tilde{\mathbf{s}}}}

\AtBeginDocument{%
  \providecommand\BibTeX{{%
    \normalfont B\kern-0.5em{\scshape i\kern-0.25em b}\kern-0.8em\TeX}}}

\setcopyright{acmcopyright}
\copyrightyear{2021}
\acmYear{2021}
\acmDOI{10.1145/1122445.1122456}

\acmPrice{15.00}
\acmISBN{978-1-4503-XXXX-X/18/06}

\begin{document}

\title{Learning to Select Historical News Articles for Interaction based Neural News Recommendation}


\author{Peitian Zhang}
\email{zpt@ruc.edu.cn}
\affiliation{%
  \institution{Renmin University of China}
  \streetaddress{}
  \city{Beijing}
  \state{}
  \country{China}
  \postcode{100872}
}
\author{Zhicheng Dou}
\email{dou@ruc.edu.cn}
\affiliation{%
  \institution{Renmin University of China}
  \streetaddress{}
  \city{Beijing}
  \state{}
  \country{China}
  \postcode{100872}
}

\author{Jing Yao}
\email{jing@ruc.edu.cn}
\affiliation{%
  \institution{Renmin University of China}
  \streetaddress{}
  \city{Beijing}
  \state{}
  \country{China}
  \postcode{100872}
}


\begin{abstract}
    The key to personalized news recommendation is to match the user’s interests with the candidate news precisely and efficiently. Most existing approaches embed user interests into a representation vector then recommend by comparing it with the candidate news vector. In such a workflow, fine-grained matching signals may be lost. Recent studies try to cover that by modeling fine-grained interactions between the candidate news and each browsed news article of the user. Despite the effectiveness improvement, these models suffer from much higher computation costs online. Consequently, it remains a tough issue to take advantage of effective interactions in an efficient way. To address this problem, we proposed an end-to-end \textbf{S}elective \textbf{F}ine-grained \textbf{I}nteraction framework (SFI) with a learning-to-select mechanism. Instead of feeding all historical news into interaction, SFI can quickly select informative historical news w.r.t. the candidate and exclude others from following computations. We empower the selection to be both sparse and automatic, which guarantees efficiency and effectiveness respectively. Extensive experiments on the publicly available dataset MIND validates the superiority of SFI over the state-of-the-art methods: with only five historical news selected, it can significantly improve the AUC by 2.17\% over the state-of-the-art interaction-based models; at the same time, it is four times faster.
\end{abstract}




\settopmatter{printacmref=false}
\maketitle

\section{Introduction}
\label{section::introduction}
Nowadays, people are overwhelmed with information, exhausted to seek things they're interested in. Online news platforms e.g. MSN News\footnote{\url{https://www.msn.com/en-us/news}} greatly alleviate this information overload problem by recommending news articles according to user's specific interests~\cite{Zheng+2018+DRN_reinforce,Wu+2019+Topic_aware,Lian+2020+Geograohy-aware,Google+2010+Bayesian_with_news_trend}. The key technology of these news platforms is personalized news recommendation~\cite{personalized_news_recommendation}. Due to the particular large-scale and time-sensitive property of news, the news recommenders must be both effective and efficient so that it can be deployed in real production systems.

\begin{figure}[t]
  \centering
  \includegraphics[width=\linewidth]{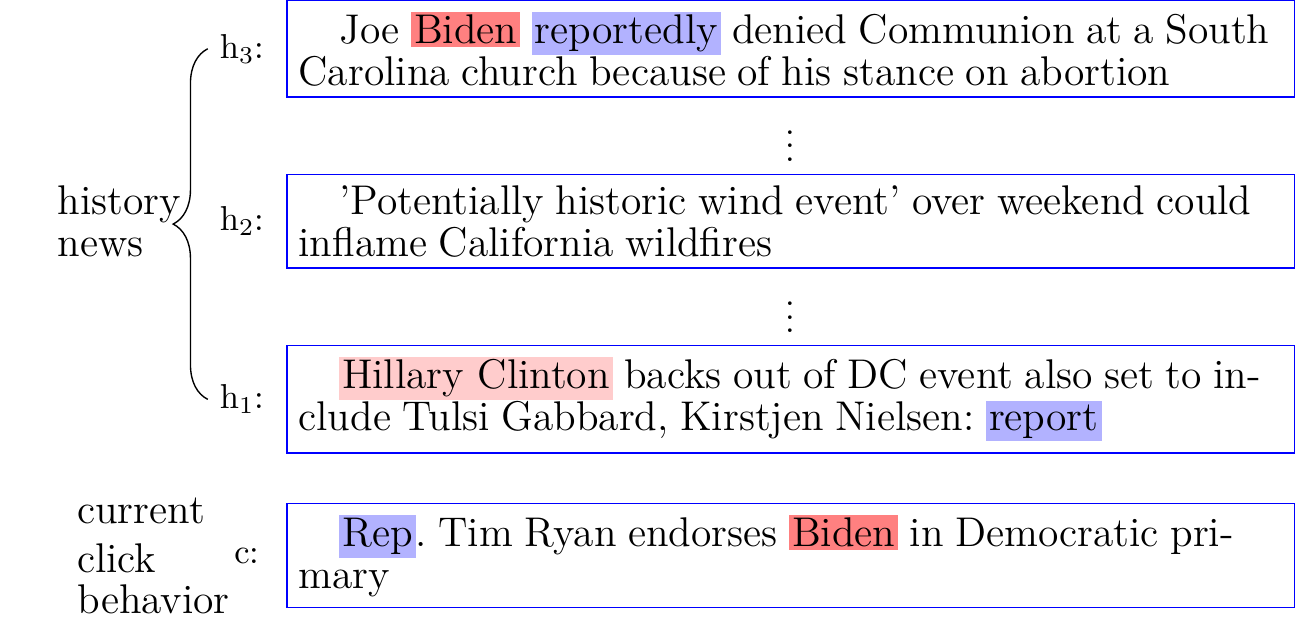}
  \caption{Example of a user's behavior log in MSN. Three historical news articles
  of the user are shown and $\mathrm{c}$ is the news she actually clicked afterward. The text marked in the same color is the fine-grained (term-term) matching signals. The darker the color, the more they match.}
  \Description{Illustration}
  \label{fig::illustration}
\end{figure}

A lot of existing news recommendation approaches~\cite{Okura+2017+GRU_user_encoder,Wu+2019+LSTUR,Yu+2019+model_long_short_term_interest_with_forget,Wu+2019+NPA,Wu+2019+NRMS,Wu+2019+NAML,Wu+2019+LSTUR,Wu+2019+Heterogeneous_user_behavior,Wu+2019+Topic_aware,Lian+2020+Geograohy-aware} follow a representation-based matching strategy. They learn a representation vector for the candidate news and encode the user's history news into a vector to form the user representation in the same semantic space. The matching score between these vectors is calculated as the click probability. However, the user vector is an aggregation of multiple historical articles, so it hardly keeps the fine-grained information and may contain noise in the articles. For example in Figure~\ref{fig::illustration}, the candidate news matches user's fine-grained interests \textit{Biden} and \textit{Hillary} (fine-grained interaction happens), which motivates the user's current click. Unfortunately, the aggregated user vector mixes all terms in $\mathbf{h}_1,\mathbf{h}_2$ and $\mathbf{h}_3$ and blurs these fine-grained interests, thus degrades the capacity of user modeling. Even worse, noises such as \textit{wind} and \textit{wildfires} are also included and they are unrelated to the current click. Though some recent ``multi-channel'' methods~\cite{Liu+2019+Hifi-ark,MemoryNetwork} attempt to cover richer information by maintaining multiple representation vectors, they are still limited to modeling fine-grained interaction in an explicit and reliable manner.

In order to capture fine-grained matching signals between the candidate news and the user, Wang et al.~\cite{Wang+2019+FIM} proposed an interaction-based model. It computes similarity matrices between the candidate news and every historical news piece of the user at word level to derive the click probability. Despite the effectiveness improvement, the model is especially slow. It has to recompute term-level interaction matrices with every historical article when scoring each candidate, which is far more expensive than dot product used in representation-based methods. Intuitively, \textbf{we shouldn't involve all the historical articles in interaction}. For example, $\mathbf{h}_2$ should be excluded within this click because it is irrelevant to the current candidate $\mathbf{c}$ and there would be no interactions between them. Tailoring the user history to several recent browsed news items seems to be a straightforward solution to save efficiency. However, \textbf{blindly interacting with only the latest browsed news limits the recommendation effectiveness}, mainly because: 1) When the length of the kept browsing history is short, there isn't sufficient news for the model to learn the user's interests well. Back to the Figure~\ref{fig::illustration}, if we cut off earlier history news $\mathrm{h}_2$ and $\mathrm{h}_3$, the interaction quality would greatly decrease since the most informative one, $\mathrm{h}_3$, is lost. 2) When the capacity becomes larger, irrelevant historical news, such as $\mathrm{h}_2$, is involved. As mentioned above, such news is harmful to the recommendation accuracy. It hardly contributes to motivate the click and would act like noise, reducing the matching score of the candidate that the user's truly interested in.

To tackle the above problem, in this paper, we propose \textbf{SFI}, a \textbf{S}elective \textbf{F}ine-grained \textbf{I}nteraction framework. The key idea of it is to \textbf{select a small number of historical news articles with higher informativeness, then perform fine-grained interaction over them only.}

The biggest challenge of SFI is to select the informative history news sparsely, precisely, and efficiently. Most previous works about feature selection employ gating operator~\cite{Review,AutoFIS,Yuan+2019+MultiHop_selector_for_chatbots}. But it cannot be directly used in SFI because gating doesn't eliminate zero entries so the total computations are not lessened. Two-stage training~\cite{Document_Cascading} is not desirable either, because no ground-truth label (indicates which historical news interacts with the candidate) is available to train the selector. Last but not least, we'd better manipulate news-level representations to guide selection for the sake of efficiency.
In this work, we design the \textbf{learning-to-select} mechanism to fulfill all our goals. Specifically, SFI learns a selection vector for each news article, and computes the cosine similarity between candidate news and every history news in the selection space, taking the result as the informativeness of each historical article. Next, we design two successive selection networks. The hard-selection network enforces sparsity. It selects $K$ most informative news and excludes others from following interactions. Within the output of the hard-selection, the following soft-selection network masks news whose informativeness is below a given threshold and attaches different weights to the unmasked ones. This refinement allows the gradient flow through to optimize the selection vectors, so that the model can learn to highlight valuable features for selection hence achieve higher effectiveness.

Extensive experiments on the publicly available dataset MIND show that SFI outperforms all baselines in terms of both effectiveness and efficiency: with only five historical news articles selected, it significantly improves the recommendation effectiveness by 2.17\% over the state-of-the-art interaction-based models with four times faster speed (almost reaches the fastest speed of representation-based methods and outperforms it by $2.71\%$ in AUC). We also comprehensively compare SFI with its naive recent $K$ counterpart and investigate the efficiency effectiveness trade-off brought by SFI.

The main contributions of this paper can be summarized into three aspects:

(1) We propose SFI, a selective fine-grained interaction framework, to take full advantage of the fine-grained interaction in a highly efficient way.

(2) We design the learning-to-select mechanism to sparsely and automatically select informative historical news w.r.t. the candidate.

(3) We conduct extensive ablation studies to verify the advantage of selection; and further investigate the efficiency-effectiveness trade-off that SFI achieves.


\section{Related Work}
\label{section::related work}
In this section, we first review the traditional recommendation methods, then the neural news recommendation methods.

\subsection{Traditional News Recommendation}
A lot of traditional recommendations methods~\cite{Das+2007+collaborative_lsh_model,Li+2011+Two_stage_hierarchical_lsh,Koren+2009+Summary_for_matrix_factorization} are based on collaborative filtering (CF). CF-based methods cluster users by ``co-visitation'' relationships to recommend news to similar users~\cite{Das+2007+collaborative_lsh_model}. Another line of CF studies apply Matrix Factorization~\cite{Koren+2009+Summary_for_matrix_factorization} and Factorization Machine~\cite{Rendle+2010+FM} to model the interaction between users and items. However, these methods face the problem of cold-start and sparsity, which is severe in news domain. They also require difficult and labor-consuming feature engineering. As the counterpart of CF, content-based recommendation methods become the main focus of news recommendation~\cite{Google+2010+Bayesian_with_news_trend, LinearUCB} because of the rich text information in news articles.

\subsection{Neural News Recommendation}
In recent years, deep learning techniques are widely used in news recommendation systems and achieve better results than traditional methods. They can be categorized as follows:

\subsubsection{Feature-based Methods}
Following traditional recommendation approaches, feature-based methods feed the model with news content together with manually designed features, then employ neural networks to model the complex interactions among all the features~\cite{Google+2016+Wide&Deep,Guo+2017+DeepFM,Lian+2018+DeepFusionModel_fm_with_attention_fusion}. For example, Cheng et al.~\cite{Google+2016+Wide&Deep} propose to combine shallow and deep neural networks to extract valuable information from a variety of manual features. Guo et al.~\cite{Guo+2017+DeepFM} add deep layers over the factorization machine to model high order interactions.

\subsubsection{Representation-based Methods}
More methods proposed to learn representations of news and users from raw texts and browsing histories respectively~\cite{Okura+2017+GRU_user_encoder,Huang+2013+DSSM,Wu+2019+NPA,Wu+2019+NRMS,Wu+2019+NAML,Wu+2019+LSTUR,Yu+2019+model_long_short_term_interest_with_forget,Wu+2019+Topic_aware,Wu+2019+Heterogeneous_user_behavior,Lian+2020+Geograohy-aware}. Numerous well designed models are proposed: multi-layer perceptron over tri-grams~\cite{Huang+2013+DSSM}, denoising auto-encoder~\cite{Okura+2017+GRU_user_encoder}, convolution neural networks ~\cite{Wu+2019+NPA,Wu+2019+NAML,Wu+2019+LSTUR,Yu+2019+model_long_short_term_interest_with_forget,Wu+2019+Topic_aware,Wu+2019+Heterogeneous_user_behavior,Lian+2020+Geograohy-aware}, and various attention-based methods~\cite{Wu+2019+NPA,Wu+2019+NRMS,Wu+2019+NAML}. Multi-channel structure is also explored~\cite{Liu+2019+Hifi-ark}.
Besides, some approaches~\cite{Wu+2019+Session_based_bipartite_graph,Hu+2020+GNN_multi_subspace_preference_disentanglement,Wu+2019+GNN_hierarchical_attention} focused on using graph neural networks to represent news and users with their neighbors. Several methods~\cite{Wang+2018+DKN,Wang+2019+Knowledge_grpah_convolution} proposed to incorporate knowledge to construct knowledge-aware representations of news and users.

In spite of the improvements these methods have made, all of them embed the news and the user into one or several one-fold vectors in the semantic space, where the fine-grained information is limited. And the representations can only meet each other in the prediction phase, which may impair fine-grained matching signals between the user and the candidate news.
\subsubsection{Interaction-based Methods}
To address the above problem, interaction-based models, which match user's interests with the candidate news at more delicate levels, are proposed.
Wang et al.~\cite{Wang+2019+FIM} designed the state-of-the-art interaction-based method for news recommendation. They
constructed segment-to-segment similarity matrices between the candidate news and every historical news article of the user from 3 different granularities. Then they use 3D-CNN to highlight salient matching signals to make recommendations. Although FIM achieves better results feature- and representation-based methods, its main drawback is the especially slow inference speed.

In this work, \textbf{we explore the interaction-based methods and aim to efficiently select fewer historical articles with higher value to perform interactions}. Several works in other fields have proposed to select important features for interaction~\cite{Review,AutoFIS,Yuan+2019+MultiHop_selector_for_chatbots}, but they all employ the gating operator, which fails to discard the unselected items hence cannot be directly used to achieve our goal. The most related work ~\cite{Document_Cascading} splits selection to another training stage, forbidding the model learning to select. Our proposed \textbf{learning-to-select} mechanism effectively addresses these issues.

\section{Our Approach}
\label{section::our approach}
\begin{figure*}[t]
  \centering
  \includegraphics[width=0.85\textwidth]{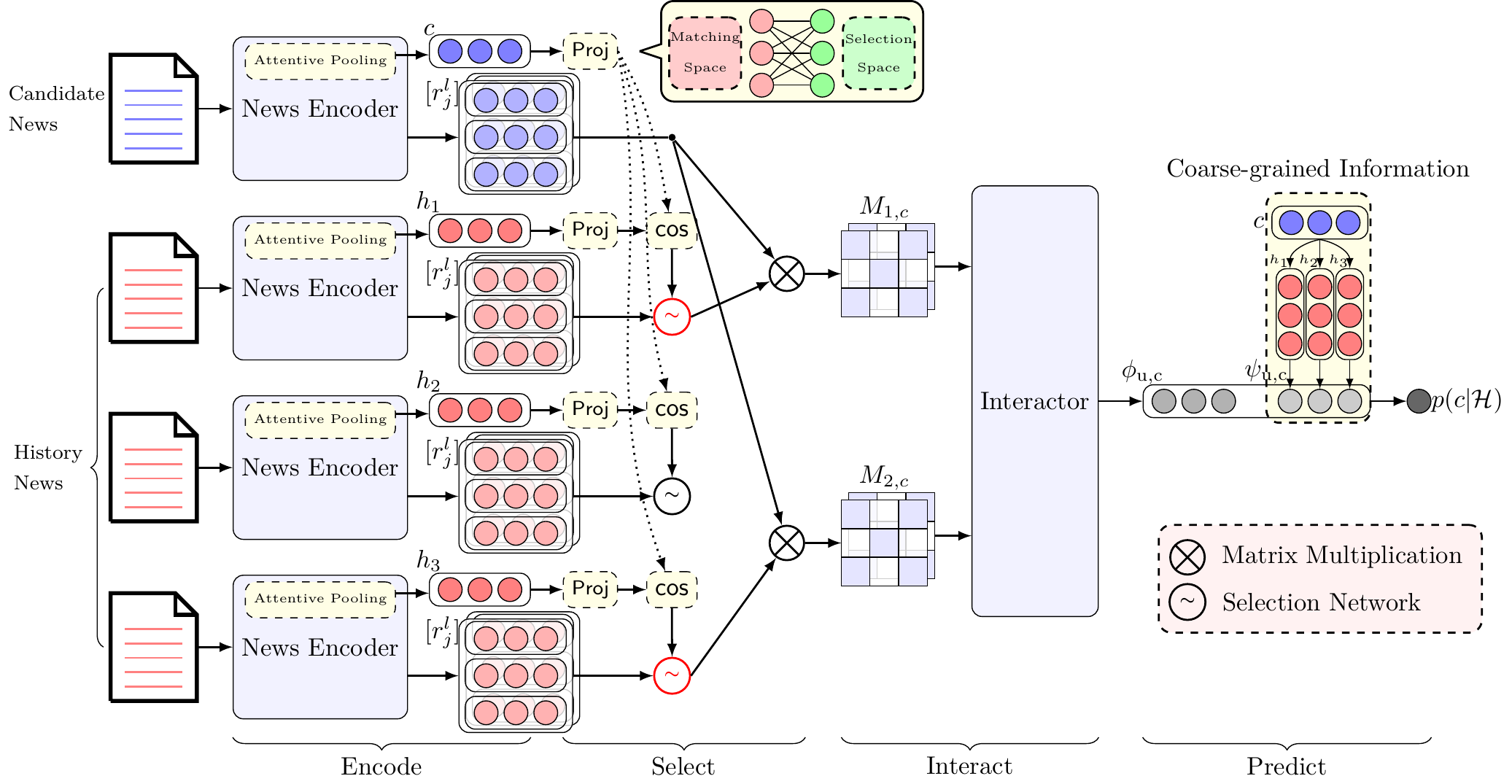}
  \caption{The architecture of our SFI model.}
  \Description{The model selects informative historical news using the efficient representation based matching, then carry out effective fine-grained term-level interaction over the selected news for final recommendation.}
  \label{fig:model}
\end{figure*}
First we formulate the news recommendation problem. Given a user $u$, we have a set of historical news articles browsed by her at the platform, denoted as $\mathcal{H} = \{\mathrm{h}_1,\mathrm{h}_2,\cdots ,\mathrm{h}_M\}$. For a candidate news $\mathrm{c}$, our goal is to infer the probability that the user clicks this news article based on her browsing history $\mathcal{H}$, denoted as $p(\mathrm{c}|\mathcal{H})$.

The architecture of SFI is presented in Figure ~\ref{fig:model}. Specifically, it contains four major modules. The \textbf{news encoder module} learns word- and news-level representations, the fine-grained ones are intended for interaction and the coarse-grained ones are further transformed for selection. The following \textbf{history selector module} manipulates news-level representation to efficiently and precisely select informative news from the user's browsing history. The fine-grained representations of selected news are fed into the \textbf{news interactor module} to compute interactions. The coarse-grained matching signals are also modeled in this module. Finally the \textbf{click predictor module} incorporates all matching signals to predict the click probability $p(\mathrm{c}|\mathcal{H})$. Next, we introduce each component in our model, especially the history selector.
\subsection{News Encoder Module}
\label{subsec::news encoder}
Since users' click decisions on news platforms are usually based on the title of news articles~\cite{Wu+2019+NPA}, the \textit{news encoder} learns the news representation from title only. Denote the word sequence of a news title as $S = \{\mathrm{w}_1,\mathrm{w}_2,\cdots,\mathrm{w}_N\}$, where $N$ is the length of $S$. First of all, we transform $S$ into a sequence of vectors $E = \{\mathbf{e}_1,\mathbf{e}_2,\cdots, \mathbf{e}_N\}$ by word embedding matrix $\mathbf{W_e}\in \mathbb{R}^{V\times D}$. $V$ is the vocabulary size and $D$ is the dimension of embeddings. Usually, the local contexts of a word across different spans play a big role in representing the word~\cite{Wu+2019+NPA,Wu+2019+NAML,Wang+2019+FIM}.
Therefore, we employ a hierarchical dilated convolution~\cite{Yu+2019+Dilated_cnn} to extract context features from different semantic granularities. In the $l$-th convolution layer, the representation of the $i$-th word is calculated as:
\begin{equation}
  r_i^l = \mathsf{ReLU}\left(\mathbf{F_w}\times \bigoplus_{k=0}^{w} \mathbf{e}_{i\pm k\delta} + \mathbf{b}\right) \in \mathbb{R}^{f_s},
\end{equation}
where $\bigoplus$ means the concatenation operation for vectors. $\mathbf{F_w}$ is the convolution kernel of size $2w + 1$, $\delta$ denotes dilation rate, $\mathbf{b}$ denotes bias and $f_s$ denotes the number of filters. A detailed description of dilated convolution can be found in~\cite{Wang+2019+FIM}. By hierarchically stacking dilated convolutions with expanding dilation rate, local contexts of different distances are fused into the word representations.
Afterward, the output of each convolution layer is appended to the final representation of $i$-th word: $\mathbf{r}_i = \{\mathbf{r}_i^l\}_{l=0}^{L}$, where $\left\{\cdot\right\}$ denotes the vertical alignment of a matrix, and $L$ denotes the total number of the stacked convolution layers.

The representation of each semantic level may contain information of different importance for matching. For example, in the news title ``Restaurants to Satisfy Late Night Cravings in Louisville and Beyond'', phrase-level local contexts ``Late Night Craving'' for the word ``Night'' matter more than that of sentence-level, e.g. ``Restaurants ... Night ... And''. Therefore, we use an attentive pooling technique~\cite{Yu+2019+model_long_short_term_interest_with_forget,Wu+2019+NAML,Wu+2019+LSTUR} to highlight the important local contexts of a single word. Specifically, a trainable vectors $\mathbf{q_l}\in \mathbb{R}^{f_s}$ is introduced as the query of attention. The representation $\mathbf{r}_i'$ of the word $\mathrm{w}_i$ that fuses information across every semantic level is computed as:
\begin{align}
  \mathbf{r}_i' = \sum_{l=1}^{L}a_{il} \mathbf{r}_i^l,\qquad \text{where }\quad a_{il} = \frac{\exp(\mathbf{q_l}^{\top} \mathbf{r}_i^{l})}{\sum_{j=1}^L \exp(\mathbf{q_l}^{\top} \mathbf{r}_j^{l})}.
  \label{eq::attentive pooling 1}
\end{align}
Similarly, different words may contribute differently in expressing the news. For example, ``Louisville'' is more informative than ``Beyond'' because it reveals the location. We use another query vector $\mathbf{q}_w\in \mathbb{R}^{f_s}$ to highlight the informative words in the news title and obtain the overall representation of the entire news:
\begin{align}
  \mathbf{r} = \sum_{i=1}^{N}a_{i} \mathbf{r}_i', \qquad \text{where }\quad a_{i} = \frac{\exp(\mathbf{q_w}^{\top} \mathbf{r}'_i)}{\sum_{j=1}^L \exp(\mathbf{q_w}^{\top} \mathbf{r}'_j)}.
  \label{eq::attentive pooling 2}
\end{align}

So far, the fine-grained representation of each word $\mathbf{r}_i = \{\mathbf{r}_i^l\}_{l=0}^{L}$ and the coarse-grained representation of the entire news $\mathbf{r}$ are generated by \emph{news encoder}. We further explore other kinds of state-of-the-art encoders, and study their performance in Section~\ref{subsec::ablation study}.

\subsection{History Selector Module}
The \textit{history selector} is the core component of our model.
It selects the informative historical news sparsely, automatically, and efficiently with a \textit{learning-to-select} mechanism. Then the selected news pieces are fed into next module for fine-grained interactions.

Denote the news-level representation of $i$-th clicked news in the user's history as $\mathbf{h}_i$,
and that of candidate news as $\mathbf{c}$. Recall that the news-level representation is attentively aggregated from the word vectors in the title, which are optimized for the final matching and thus aren't selection-oriented. Therefore, directly using $\mathbf{h}_i$ and $\mathbf{c}$ for selection may lead to sub-optimal results. In learning-to-select, we project all the news-level representations into the same selection space using a fully-connected network to mitigate such conflicts:
\begin{align}
  \mathsf{Proj}(\mathbf{r}) &= \mathbf{W_p}\mathbf{r} + \mathbf{b},
\end{align}
where $\mathbf{r}$ could be either $\mathbf{h}_i$ or $\mathbf{c}$.
Considering selection efficiency, we define the candidate-aware informativeness of every historical article as the cosine similarity between selection vectors:
\begin{align}
  \mathbf{s} &= \{  \mathrm{cos}\left( \mathsf{Proj}(\mathbf{h}_i), \mathsf{Proj}(\mathbf{c})\right)\}_{i=1}^M.
\end{align}

In selection, no supervision signals are available for $
\mathbf{s}$, so it's critical to optimize the parameters end-to-end to allow the model learn valuable features for selection. Besides, sparsity must be enforced, otherwise the model would still be slow. Keeping both constraints in mind, we design two complementary selection networks.

\subsubsection{Hard-Selection Network}
This sub-module enforces sparsity: it keeps the top $K$ most informative history news and discards others. Formally,
\begin{align}
  \mathbf{x} &= \mathsf{argTOPK}(\mathbf{s}) \in \mathbb{R}^{K},
\end{align}
where $\mathsf{argTOPK}(\mathbf{s})$
gets the index of the top $K$ value in the vector $\mathbf{s}$.
The corresponding history news sliced by $\mathbf{x}$ is selected and its fine-grained representations will get involved in the fine-grained interaction later. To do that, we first transform $\mathbf{x}$ to one-hot encoding matrix $\mathbf{X} = \mathsf{one\_hot}(\mathbf{x})\in \mathbb{R}^{K\times M}$, where the $i$-th row in $\mathbf{X}$ is the one-hot vector of $\mathbf{x}_i$, then use matrix multiplication to prune the browsing history to a smaller size:
\begin{align}
  \hat{H} = \mathbf{X}\otimes H,
  \qquad \hat{\mathbf{s}} = \mathbf{X}\otimes \mathbf{s}.
\end{align}
where $H = \{\{\mathbf{h}_i^l\}_{l=0}^L\}_{i=1}^M\in \mathbb{R}^{M\times L \times f_s}$ is the fine-grained representation tensor
of user's browsed news, $\hat{H}\in \mathbb{R}^{K\times L \times f_s}$ is that of the selected news.
By regulating hyper parameter $K$, we can elastically control the model's efficiency.

However, this sub-module has three defects that may limit the effectiveness: 1) Because $\mathbf{X}$ is sparse, the gradient cannot be passed to optimize $\mathbf{W_p}$; 2)  Because the informativeness distributions of different users vary greatly, some noisy news articles are not filtered out among top $K$; 3) All of the selected news articles are weighted equally even though some of them are more informative. To tackle the above problems, a soft refinement is proposed.

\subsubsection{Soft-Selection Network}
This sub-module makes the gradient flow through selection. It is essentially a gating operator with  a threshold~\cite{Yuan+2019+MultiHop_selector_for_chatbots}, which further rules out noise (namely authentically uninformative news) and improves effectiveness. Given the output of hard-selection, the soft-selection network masks the news whose informativeness is below the threshold and attaches different importance to the unmasked ones:

\begin{align}
     \tilde{H} &= \hat{H} \odot \mathsf{Expand}(\mathbf{\tilde{s}}), \qquad\mathbf{\tilde{s}} = \sigma(\mathbf{\hat{s}}),\\
    \sigma(s_i) &= \begin{cases}
    0 & s_i < \gamma,\\
    h_i & \mathrm{otherwise}.
  \end{cases}
\end{align}
$\odot$ is Hadamard Product, and $\gamma$ denotes the threshold. $\sigma(\cdot)$ is element-wise.
$\mathsf{Expand}(\tilde{\mathbf{s}})$ repeats the elements in $\tilde{\mathbf{s}}$, expanding it into $\mathbb{R}^{K\times L\times f_s}$. $\tilde{H}$ is the refined $\hat{H}$, where all representations of the news whose informativeness is lower than $\gamma$ are masked as $0$.

The number of the authentically informative news articles is floating per candidate, so a dynamic quantity of news items is kept. This entitles more flexibility to the selection operation. Meanwhile, with $\tilde{s}$ attached to $\hat{H}$, the \textit{news interactor} can attend to more informative ones, and the parameters in $\mathsf{Proj}(\cdot)$ can be optimized by the selecting step since the element-wise multiplication is differentiable. This helps SFI to learn features that are important for selection and will enhance the effectiveness remarkably.

In back propagation, gradient from the loss function is applied to the \textit{news interactor}, then to the selected fine-grained representation tensor $\tilde{H}$. For simplification, $\tilde{H}$ is reshaped into a vector $\tilde{\mathbf{R}}\in \mathbb{R}^{1\times(K\times d)}$ where $d=L\times f_s$, together with its gradient $\nabla_{\tilde{\mathbf{R}}} = \mathsf{reshape}(\nabla_{\tilde{H}})\in \mathbb{R}^{1\times(K\times d)}$. The same operation is taken for $\hat{H}$, forming $\hat{\mathbf{R}}\in \mathbb{R}^{1\times(K\times d)}$. Then the gradient for $\mathbf{W_p}$ is:
\begin{align}
\label{eq::gradient}
\dw &= \mathbf{Z}^{\top}\otimes\dz\otimes\mathbf{c}_1' + \mathbf{c}^{\top}\otimes \dz^{\top}\otimes\mathbf{Z}_1,\\
\dz &= \mathbf{X}^\top\otimes\dzz \notag\\
    &= \mathbf{X}^\top\otimes\left(\dfz\odot g(\hat{\mathbf{s}})\right) \notag\\
    &= \mathbf{X}^\top\otimes\left(\left\{
        \sum_{j=1}^{d}\left(\dR\otimes \mathsf{Diag}(\hat{\mathbf{R}})\right)[i,j]
        \right\}_{i=1}^{K}\odot g(\hat{\mathbf{s}})\right),
\end{align}
where $\mathsf{Diag}(\hat{\mathbf{R}})$ is the matrix with the elements of $\hat{\mathbf{R}}$ as the diagonal. $\mathbf{Z} = \{\mathbf{h}_i\}_{i=1}^M\in \mathbb{R}^{M\times f_s}$ is the news-level representation matrix of the historical news, and $\mathbf{Z}_1 = \{\mathsf{Proj}(\mathbf{h}_i)\}_{i=1}^M, \mathbf{c}_1 = \mathsf{Proj}(\mathbf{c})$ are the corresponding selection vectors. $g(s_i)$ is the derivative for $\sigma(s_i)$:
\begin{equation*}
    g(s_i) = \begin{cases}
    0&s_i < \gamma,\\
    1&\mathrm{otherwise}.
\end{cases}
\end{equation*}
In this way, the gradient safely flows through the selection stage and reaches $\mathbf{s}$, to increase the score of the useful news pieces and vice versa. It is further spread to optimize $\mathbf{W_p}$ to achieve the above adjustment, however, only from the selected entries.

\subsection{News Interactor Module}
The selected historical news articles are fed into this module to perform fine-grained interactions with the current candidate. We denote the representation of the words in $v$-th selected news as $\mathbf{d}_v = \{\mathbf{t}_i\}_{i=1}^N$ where $\mathbf{t}_j = \{\mathbf{t}_j^l\}_{l=0}^L\in \mathbb{R}^{L\times f_s}$ is the stacked representation of $j$-th word. Similarly, the representation of each word in the current candidate news is $\mathbf{c}^\mathrm{f} = \{\mathbf{p}_j\}_{j=1}^{N}$. Resembling FIM~\cite{Wang+2019+FIM}, we construct pair-to-pair similarity matrix $\mathbf{M}^l_{v,\mathrm{c}}$ of $l$-th semantic granularity, where each entry is the scaled dot product between the fine-grained representations of $v$-th selected news and the candidate news:
\begin{equation}
  \mathbf{M}_{v,\mathrm{c}}^l[i,j] = \frac{\mathbf{t}_i[l]^{\top} \mathbf{p}_j[l]}{\sqrt{f_s}} \in \mathbb{R}^{N\times N}.
\end{equation}

Next, the similarity matrices of each granularity across all the selected history news are fused into a 3D cube $O\in \mathbb{R}^{L\times K \times N\times N}$, where a series of 3D CNN and 3D max pooling is applied to highlight the significant matching signals. Outputs of the final pooling layer are flattened as the vector containing fine-grained interactive information across the user and candidate news, denoted as $\phi_{\mathrm{u},\mathrm{c}}$. Other state-of-the-art interactors are studied in Section\ref{subsec::ablation study}.

In SFI, fine-grained matching information $\phi_{\mathrm{u},\mathrm{c}}$ only engage selected news articles. However, it is important not to leave out the unselected ones. Although conducting fine-grained interactions on them is unnecessary, we still value the coarse-grained matching signals of them, which come from the matching between news-level representations:
\begin{align}
  \psi_{\mathrm{u},\mathrm{c}} = \{\psi_{\mathbf{h}_1,\mathbf{c}},\psi_{\mathbf{h_2},\mathbf{c}},\cdots,\psi_{\mathbf{h}_M,\mathbf{c}}\}, \qquad
  \psi_{\mathbf{h}_i,\mathbf{c}} = \mathbf{h}_i^{\top}\mathbf{c}.
\end{align}
$\psi_{\mathrm{u},\mathrm{c}}$ gives an overall matching degree of the user and the candidate news and is complementary to $\phi_{\mathrm{u},\mathrm{c}}$. It facilitates the model to learn more precise correspondences between the matching signals and the click probability. Another critical point is that by involving $\psi_{\mathrm{\mathrm{u},\mathrm{c}}}$ to score the candidate, the gradient can be delivered by all of the historical news articles rather than only the selected ones.

\subsection{Click Predictor}
\label{subsec::click predictor}
The \emph{click predictor} module incorporates the output from \textit{news interactor} then predicts the probability of a user clicking on a candidate news article. The news articles with higher click probability are ranked higher in the final user interface.

Given vectors containing course- and fine-grained matching information,  $\psi_{\mathrm{u},\mathrm{c}}$ and $\phi_{\mathrm{u},\mathrm{c}}$ respectively, we propose to incorporate both by:
\begin{equation}
  y_{\mathrm{u},\mathrm{c}} = \mathbf{W_c}\left\{ \phi_{\mathrm{u},\mathrm{c}},\psi_{\mathrm{u},\mathrm{c}} \right\} + \mathbf{b}.
\end{equation}

Following ~\cite{Huang+2013+DSSM, Wu+2019+NPA}, we use negative sampling to simulate the unbalanced distribution of clicked news in an impression. For each ground-truth candidate, we randomly sample $m$ news that is not clicked by her in the same impression as negative samples:
\begin{equation}
  \hat{p}(\mathrm{c}|\mathcal{H}) = \frac{\exp(y_{\mathrm{u},\mathrm{c}}^+)}{\exp(y_{\mathrm{u},\mathrm{c}}^+) + \sum_{j = 1}^m\exp(y_{\mathrm{u},\mathrm{c}}^-)}.
\end{equation}
Thus, it is converted to a $m+1$ classification problem, and the negative log likelihood loss is going to be minimized when training:
\begin{equation}
  \mathcal{L} = -\sum_{\mathrm{c}\in\mathcal{S}} \log\hat{p}(\mathrm{c}|\mathcal{H}),
\end{equation}
where $\mathbf{c}$ is the ground-truth news piece which the user clicked, and $\mathcal{S}$ denotes all training samples.

Finally, we jointly train the \emph{news encoder}, \emph{history selector}, \emph{news-interactor} and \emph{click predictor} through the final click signal. In such a way, the model can better learn dependencies among modules.

\section{Experimental}
\label{section::experiment}
\subsection{Datasets and Experimental Settings}
\begin{table}[t]
  \caption{Statistics of the MIND dataset}
  \label{table::mind statistics}
  \begin{tabular}{p{0.23\linewidth}p{0.15\linewidth}|p{0.23\linewidth}p{0.15\linewidth}}
    \toprule
    \textbf{\#users}&1,000,000&\textbf{\#news}&161,013\\
    \hline
    \textbf{\#impressions}&15,777,377&\textbf{\#clicks}&24,155,470\\
    \hline
    \textbf{avg. title len}&11.52&\textbf{avg. his len}&32.99\\
    \bottomrule
  \end{tabular}
\end{table}
Our experiments are conducted on MIND~\cite{Wu+2020+MIND}, a large-scale dataset collected from the users' click logs of the Microsoft News platform from Oct. 12 to Nov. 22, 2019. The statistics of MIND are shown in Table~\ref{table::mind statistics}. We use the same training-testing partition as~\cite{Wu+2020+MIND}.

In our experiments, the dimension $D$ of word embeddings is set to $300$. We use the pre-trained Glove embeddings~\cite{Pennington+2014+Glove}, to initialize the embedding matrix $\mathbf{W_e}$. The maximum length of news titles is set at $20$.  the maximum number of clicked news for learning user representations was set to $50$. In the \emph{news encoder}, we stack $3$ convolution layers with dilation rates $[1,2,3]$. The kernel size and the number of filters is set to $3$ and $150$ respectively. We employ a 2-layer composition for \emph{news-interactor} module, the output channels and the window size is set at $32-[3,3,3]$ and $16-[3,3,3]$. Each convolution component is followed by a max pooling layer with size $[3,3,3]$ and stride $[3,3,3]$. We apply the dropout strategy ~\cite{Srivastava+2014+Dropout} to the word embedding layer to mitigate overfitting. The dropout rate is set at $0.2$. Adam~\cite{Diederik+2015+Adam} is used as the optimization algorithm.

The batch size is set to $100$ when training and $400$ when predicting, and the encoding process is executed offline when predicting. Since there are $40$ kernels in total on our machine, we set $40$ parallel threads to load data in order to minimize the latency caused by processing data. We independently repeat each experiment for $5$ times and report the average performance. We conduct all experiments on a machine with Xeon(R) Silver 4114 CPUs and a TITAN V GPU\footnote{We will release the code and scripts based upon the acceptance of the paper.}.

\subsection{Evaluation Metrics}
Following existing studies, we use the average AUC, MRR, nDCG@5, and nDCG@10 scores over all impressions to evaluate the effectiveness of the models. All results come from the official test entry. Moreover, given the same batch size, we use the prediction speed i.e. iterations per second to evaluate the efficiency. In one iteration, the batch size of candidate news articles is scored.

\subsection{Baselines}
We compare SFI with the following baseline methods:

(1) General Recommendation Methods: \textbf{LibFM}~\cite{Rendle+2012+libFM}, a state-of-the-art feature-based matrix factorization approach for recommendation\footnote{The TF-IDF features are used.}; \textbf{DSSM}~\cite{Huang+2013+DSSM}, a deep structured semantic model that uses multiple dense layers upon tri-grams. All of the users' clicked news are concatenated as the query, and the candidate news is regarded as documents; \textbf{Wide\&Deep}~\cite{Google+2016+Wide&Deep}, a widely used recommendation method that uses the combination of a wide channel and a deep channel for memorization and generalization; \textbf{DeepFM}~\cite{Guo+2017+DeepFM}, a popular neural recommendation method which combines factorization machine with deep neural networks;

(2) Representation-based Methods:
\textbf{DFM}~\cite{Lian+2018+DeepFusionModel_fm_with_attention_fusion}, which uses dense layers for different channels and attentively fuse outputs;
\textbf{GRU}~\cite{Okura+2017+GRU_user_encoder}, which learn news representations with an auto-encoder and utilizes GRU to learn user representations;
\textbf{Hi-Fi Ark}~\cite{Liu+2019+Hifi-ark}, a multi-channel representation approach for recommendation;
\textbf{NPA}~\cite{Wu+2019+NPA}, which highlights informative words and news with personalized attention;
\textbf{NRMS}~\cite{Wu+2019+NRMS}, which learns delicate representations of news and users by multi-head self-attention;
\textbf{LSTUR}~\cite{Wu+2019+LSTUR}, which models long- and short-term user interests with GRU;

(3) Interaction-based Methods:
\textbf{FIM}~\cite{Wang+2019+FIM}, the state-of-the-art interaction-based approach for neural news recommendation, which encodes news by hierarchical dilated CNN and performs interaction between each of the user browsed news articles and the candidate.
\textbf{Recent($K$)}, the naive counterpart of SFI, which keeps the recent $K$ historical news for interaction only (Recent($50$) equals FIM).

\section{Experimental Results and Analysis}
\subsection{Overall Results of Effectiveness}
\label{subsec::experimental result}
\begin{table}[t]
  \centering
  \caption{The performance of different methods for news recommendation. The number of news items involving in interactions is bolded in ( ) for interaction-based methods. The result with superscript * is referencing the one in ~\cite{Wu+2020+MIND} where MIND is presented. $\dagger$ indicates a significant improvement over all baselines with paired t-test ($p<0.01$).}
  \label{table::performance}
  \small
  \begin{tabular}{llcccp{0.1\linewidth}}
    \toprule
    \textbf{Type} & \textbf{Methods}&\textbf{AUC}&\textbf{MRR}&\textbf{NDCG@5}&\textbf{NDCG@10}\\
    \midrule
    \multirow{4}{0.12\linewidth}{General methods} &
    LibFM* & $59.93$ &$0.2823$&$30.05$&$35.74$\\
    &DSSM* &$64.31$&$0.3047$&$33.86$&$38.61$\\
    &Wide\&Deep* & $62.16$&$0.2931$&$31.38$&$37.12$\\
    &DeepFM* & $60.30$ & $0.2819$&$30.02$&$35.71$\\
    \midrule
    \multirow{5}{0.12\linewidth}{Represent. based} &
    DFM* & $62.28$ & $0.2942$ & $31.52$ & $37.22$\\
    & GRU* & $65.42$ & $0.3124$ & $33.76$ & $39.42$\\
    & Hi-Fi Ark & $65.87$ & $0.3119$ & $33.64$ & $39.25$\\
    & NPA* & $66.69$ & $0.3224$ & $34.98$ & $40.68$\\
    & LSTUR* & $67.73$ & $0.3277$ & $35.59$ & $41.34$\\
    & NRMS* & $67.76$ & $0.3305$ & $35.94$ & $41.63$\\
    \midrule
    \multirow{4}{0.12\linewidth}{Interaction based} &
    FIM & $68.12$ & $0.3354$ & $36.45$ & $42.11$\\
    \cmidrule{2-6}
    & Recent ($\mathbf{5}$) & $65.39$ & $0.3164$ & $34.14$ & $39.78$\\
    & SFI ($\mathbf{5}$) & $69.60^\dagger$ & $0.3475^\dagger$ & ${37.86}^\dagger$ & ${43.51}^\dagger$\\
    & SFI ($\mathbf{50}$) & $\mathbf{69.95}^\dagger$ & $\mathbf{0.3503}^\dagger$ & $\mathbf{38.31}^\dagger$ & $\mathbf{43.97}^\dagger$\\
    \bottomrule
  \end{tabular}
\end{table}

The overall recommendation effectiveness of all models is shown in Table~\ref{table::performance}. Based on the results, we have the following observations:

(1) \textbf{Our proposed model SFI consistently outperforms other baselines in terms of all metrics}. On the one hand, SFI captures fine-grained interactions to model user interests, gaining $2.71\%$ up to $3.23\%$ AUC improvements over all of the state-of-the-art representation-based methods. On the other hand, SFI$(K)$ outperform Recent($K$) baseline by $6.4\%$ and $2.17\%$ when $K=5$ and $50$ respectively. This result substantiates the power of the \textit{learning-to-select} mechanism.

(2) The variant SFI($50$) that keeps the entire browsing history outperforms SFI($5$) that selects only $5$ historical news. This is as expected because after removing noise, SFI($50$) covers richer information to model the user. Interestingly, the improvement is tiny compared with the margin between Recent($50$) i.e. FIM and Recent($5$). We will study this phenomenon in detail in Section~\ref{subsec::efficency-effectiveness trade-off}.

(3) The interaction-based methods for news recommendation outperform all representation-based methods, which validates the benefit of capturing fine-grained matching signals. However, simply pruning the user's history to a smaller size to save speed is not feasible because it hurts the effectiveness seriously.

Without Bert~\cite{Bert}, expanded SFI (with extra news abstract) ranks among the top $15$ on the official testing leaderboard.

\subsection{Results of Efficiency}
Since the motivation of SFI is mainly concerned with efficiency, we further compare the inference speed between SFI and several baselines. Results in Table~\ref{table::computational cost} substantiates the superiority of SFI: with the selection capacity of $\mathbf{5}$, it can infer almost \textbf{four} times faster than the state-of-the-art interaction-based method, while significantly improving the recommendation effectiveness. SFI($5$) also achieves comparable speed with the state-of-the-art representation-based method NRMS, and outperforms it by $\mathbf{2.71}\%$ in AUC.
The efficiency of SFI($K$) and Recent($K$) significantly drops from $K=5$ to $K=25$, which will be further studied in Section~\ref{subsec::efficency-effectiveness trade-off}.

\begin{table}[t]
  \centering
  \caption{The inference speed comparison of different methods for news recommendation. The improvement over FIM is given in the bracket.}
  \label{table::computational cost}
  \small
  \begin{tabular}{p{0.2\linewidth}p{0.25\linewidth}cc}
    \toprule
    \textbf{Methods}&\textbf{Inference Speed}&\textbf{AUC} & \textbf{nDCG@5}\\
    \midrule
    NRMS & $121.54$ & $67.76$ & $35.94$\\
    FIM & $20.53$ & $68.12$ & $36.45$\\
    \midrule
    Recent ($\mathbf{5}$) & $125.96(\uparrow 517\%)$ & $65.39(\downarrow 4.01\%)$ & $34.14(\downarrow 6.34\%)$\\
    Recent ($\mathbf{25}$) & $40.05(\uparrow 96\%)$ & $67.32(\downarrow 1.17\%)$ & $35.16(\downarrow 3.54\%)$\\
    \midrule
    SFI ($\mathbf{5}$) & $\mathbf{99.57(\uparrow 385\%)}$ & $69.60(\uparrow 2.17\%)$ & $37.86(\uparrow 3.87\%)$\\
    SFI ($\mathbf{25}$) & $33.11(\uparrow 66\%)$ & $69.75(\uparrow 2.39\%)$ & $38.01(\uparrow 4.28\%)$\\
    \bottomrule
  \end{tabular}
\end{table}

\subsection{Ablation Study}
\label{subsec::ablation study}

Since SFI is essentially a flexible framework, we conduct extensive ablation studies to gain comprehensive insights into every module. In each subsection, we pose our claim first before explanations.

\subsubsection{\textbf{HDCNN and 3DCNN are the most effective encoder and interactor among a variety of state-of-the-art architectures.}}
\begin{table}[t]
    \centering
    \caption{The effectiveness of SFI with different \textit{news encoders} and \textit{news interactors}.}
    \small
    \begin{tabular}{c|c|c|c|c}
    \specialrule{.1em}{.05em}{.05em}
    \diagbox[width=10em]{\textbf{Encoders}}{\textbf{Interactors}}&\textbf{2D-CNN}&\textbf{3D-CNN}&\textbf{MHAI}&\textbf{KNRM}\\
    \hline
    \textbf{PCNN} & $63.71$ &$63.95$&$65.54$&$63.24$\\
    \textbf{HDCNN}& $68.45$ &$\mathbf{69.56}$&$63.66$&$60.01$\\
    \textbf{MHA} & $66.44$ &$65.56$&$61.31$&$63.92$\\
    \textbf{LSTM} & $68.58$ &$67.91$&$68.39$&$64.12$\\
    \bottomrule
    \end{tabular}
    \label{table::encoders and interactors}
\end{table}
In the interaction-based workflow, the history selector can be easily inserted between any kind of news encoder and interactor. This flexibility motivates us to study how state-of-the-art encoders and interactors would perform. We compare among 1D-CNN with Personalized Attention~\cite{Wu+2019+NPA} (denoted as PCNN), Hierarchical Dilated CNN~\cite{Wang+2019+FIM} (denoted as HDCNN), Multi-head Self Attention~\cite{Wu+2019+NRMS} (denoted as MHA), LSTM for the \textit{news encoder} and 2D-CNN, 3D-CNN~\cite{Wang+2019+FIM}, KNRM~\cite{Xiong+2017+KNRM}, Multi-Head Self Attention~\cite{Wu+2019+NRMS} (denoted as MHAI) for the \textit{news interactor}. The AUC scores are reported in Table~\ref{table::encoders and interactors}. We find \textbf{Hiearchical Dilated CNN} combined with \textbf{3D-CNN} is the best setting.

\subsubsection{\textbf{Every sub-module in history selector is critical to improving effectiveness.}}
\begin{figure}[t]
    \centering
    \includegraphics[width=\linewidth]{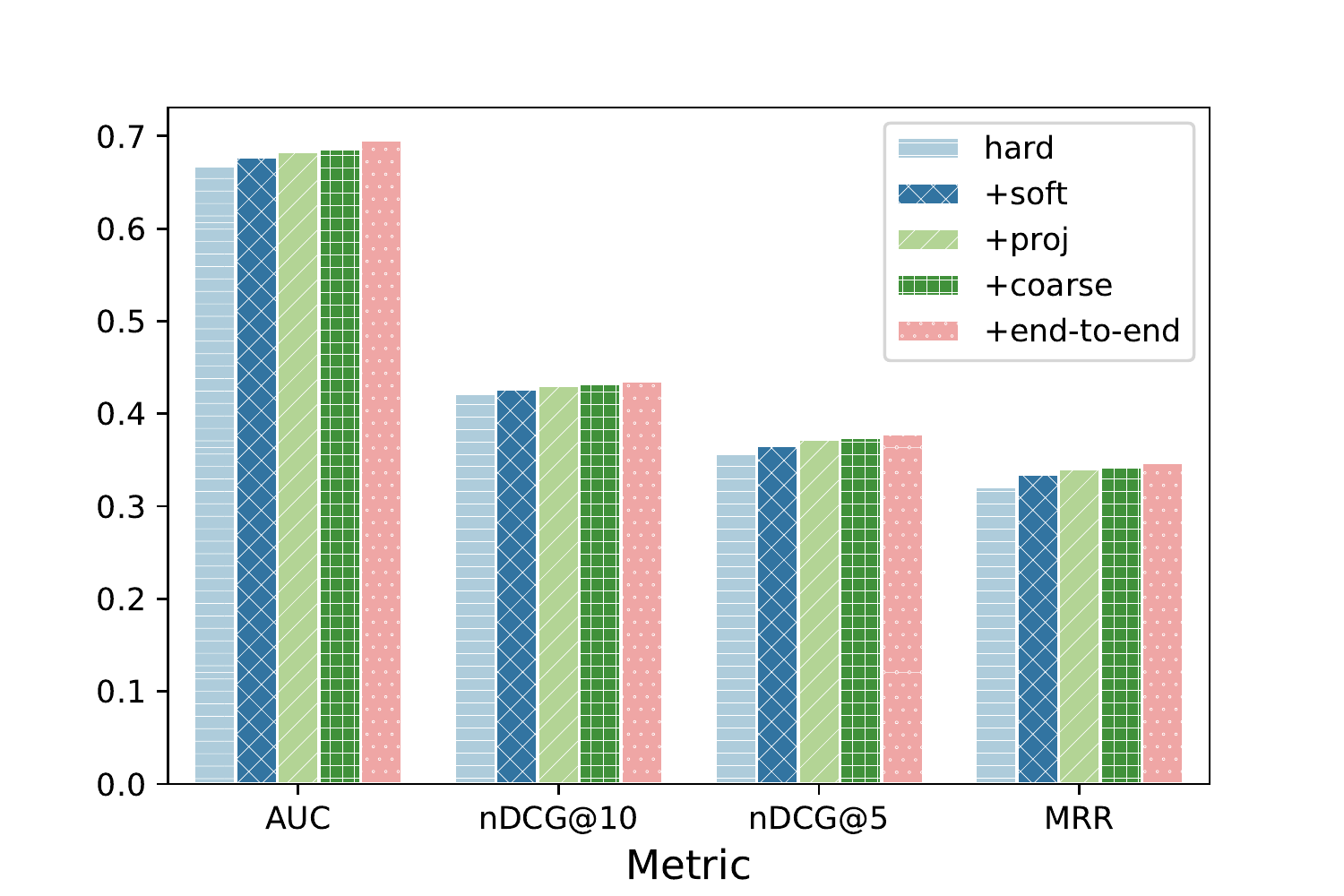}
    \caption{The effectiveness of history selector and coarse-grained information.}
    \label{fig::selection network and coarse-grained}
\end{figure}

The \textit{learning-to-select} mechanism comprises three parts: a selection projection, a hard selection, and a soft refinement. The hard selection is the cornerstone of our work so we no longer verify its impact. For the other two components, we compare SFI with the variant that applies only hard-selection, and that applies hard-selection followed by soft-selection without learning extra selection vectors. The result is reported in Figure~\ref{fig::selection network and coarse-grained}. As we observe, the soft-selection network improves the effectiveness. This is because it filters the authentically uninformative history news, and makes the gradient flow through to optimize the representation vectors used for selection. However, without selection projection, these news-level representations are optimized for two incompatible goals: selecting and matching, which may decrease the recommendation accuracy. Experiments validates our claim: the model benefits a lot from selection projection. Thanks to it, SFI can encode selective features into selection vectors, leaving the news-level representations to focus on the final matching.

\subsubsection{\textbf{The coarse matching signals of the unselected articles are also important.}}
In Figure~\ref{fig::selection network and coarse-grained}, SFI outperforms its variant that totally abandons the coarse-grained matching signals.  This observation verifies that the coarse-grained matching signals are complementary. Note that doing so won't reduce efficiency because batched matrix multiplication is fast on GPU.

\subsubsection{\textbf{SFI benefits from end-to-end training.}}
When deployed in production, the news encoding process could be done offline to speed up inference, known as a pipeline convention. It's natural to migrate it to the training phase, where we first pre-train SFI without the \emph{history selector} to acquire coarse-and-fine representations of news. Then we replace the \emph{news encoder} with a lookup table constructed from these representations and fine-tune them with \emph{history selector} applied. End-to-end training is the counterpart, in which we jointly train the \emph{news encoder}, \emph{history filter}, \emph{news-interactor} and \emph{click predictor} by the final classification loss simultaneously. In Figure~\ref{fig::selection network and coarse-grained}, the first four bars in every group are the performance of SFI trained in pipeline, and the last bar is that of trained end-to-end under the same setting. As expected, end-to-end training leads to better effectiveness. It's because optimizing the parameters rather than directly updating the vectors can make the \textit{news encoder} learn more precise representations for both selection and interaction. Also, the modules can better learn the correspondence among them in end-to-end training.

\subsection{Hyper Parameter Analysis}
\label{subsec::hparam analysis}
\begin{figure}
    \includegraphics[width=\linewidth]{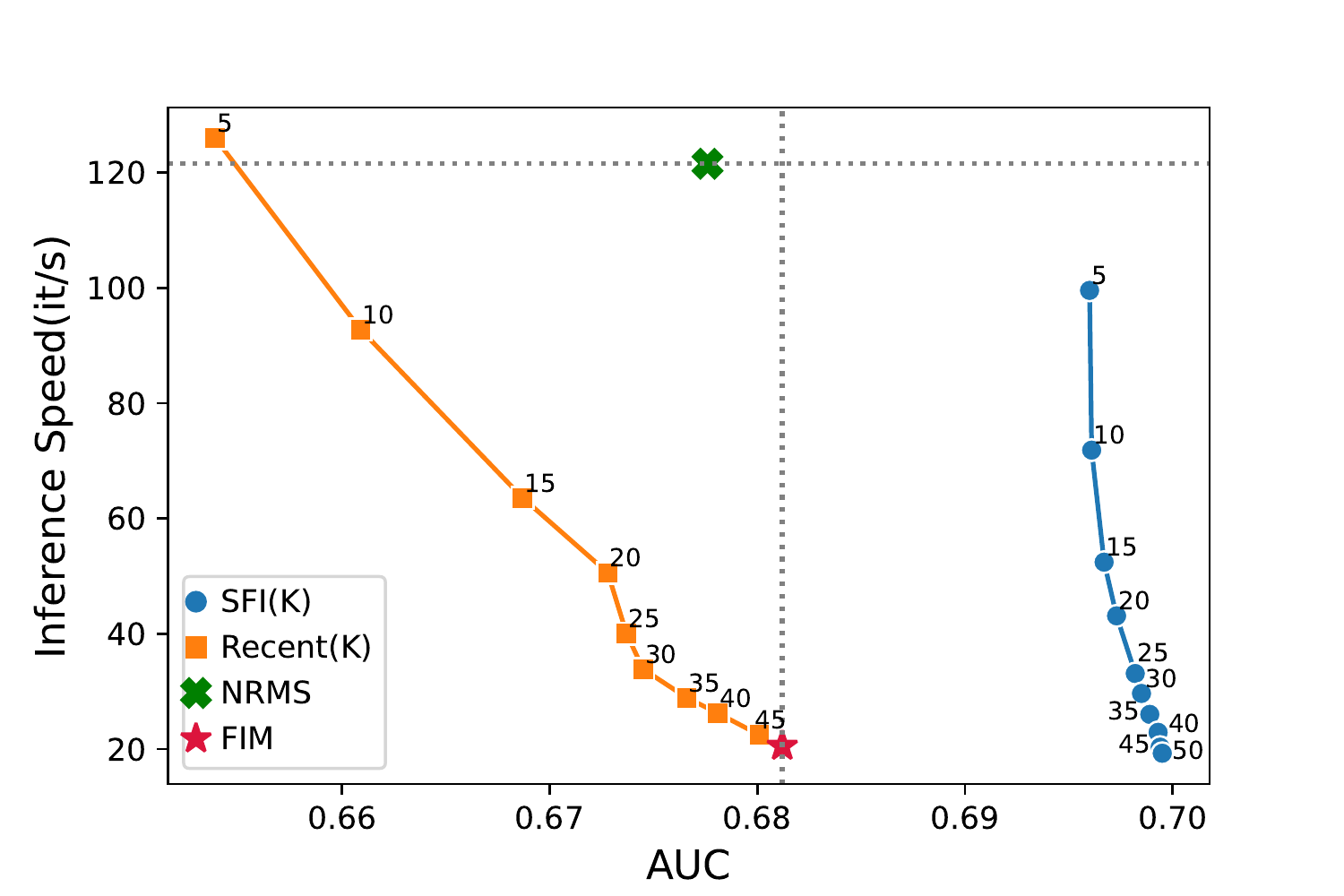}
    \caption{The efficiency and effectiveness of SFI with different numbers of selected news. The number next to the marker indicates the selected news count.}
    \label{fig::efficiency effectiveness tradeoff}
\end{figure}

\begin{figure}[t]
    \includegraphics[width=\linewidth]{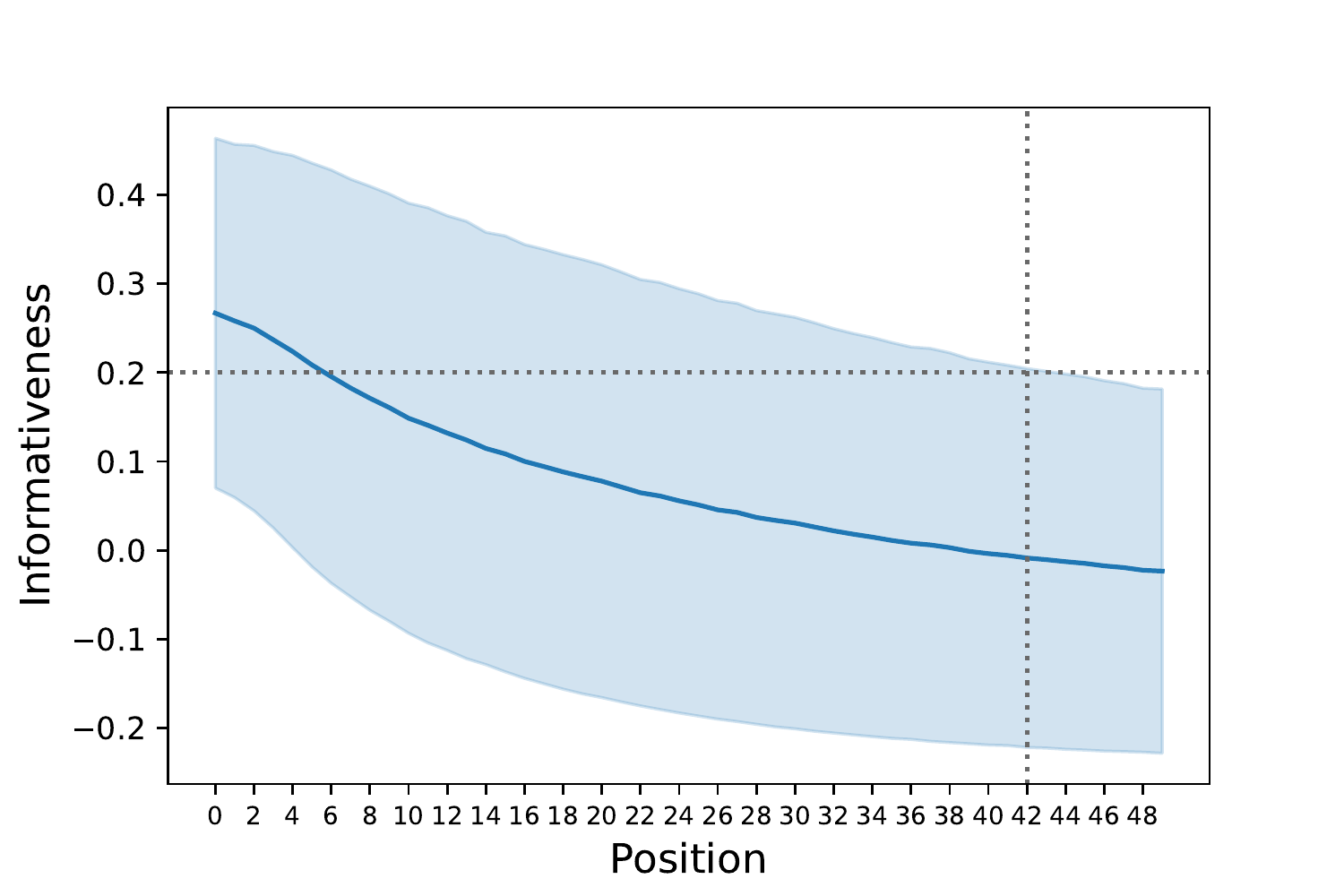}
    \caption{The informativeness score of history news at each position. Smaller x-axis represents more recent history.}
    \label{fig::informativeness
    distribution}
\end{figure}

\subsubsection{Influence of the Interaction Capacity}
\label{subsec::efficency-effectiveness trade-off}
The predefined interaction capacity $K$ is the most important hyper parameter since the efficiency-effectiveness trade-off is up to it. We study its influence by drawing the inference speed curve against the AUC score of the model with different $K$ in Figure~\ref{fig::efficiency effectiveness tradeoff}. Motivated by several observations in Section~\ref{section::experiment}, we also include Recent($K$), which only interacts with the latest $K$ historical news to save efficiency.
We add two dashed lines to mark the best effectiveness and efficiency that baseline models ever achieve. The optimal result would be situated at the upper right corner.

According to the figure, we find:
\textbf{First}, from $K=5$ to $K=50$, SFI($K$) is far more effective than its naive counterpart Recent($K$), this again validates the effectiveness of \textit{history selector}. However, due to the time consumption of selection, SFI($K$) is a bit slower. Overall, SFI($5$) is the optimal setting because it greatly outperforms NRMS and all Recent($K$) including FIM, while providing a much higher efficiency over FIM.
\textbf{Second}, when $K$ is growing, the effectiveness of both Recent($K$) and SFI($K$) is improving, which is because a bigger capacity keeps richer information to learn the user's interests. As a side effect, the model becomes slower.
\textbf{Third}, as $K$ increases, the effectiveness of SFI($K$) grows slower than Recent($K$) and is about saturated at $K=40$. Intuitively, with the \textit{learning-to-select} mechanism, SFI($K$) can consistently select the most effective articles for interaction, so increasing the capacity only brings a little more valuable information. In contrast, Recent($K$) cannot access the informative historical news in earlier history unless the capacity is big enough.

These observations motivate us to study what informativeness scores of the historical articles at different positions are learnt by the model itself. We report the informativeness score at each history position (averaged from $K=5$ to $K=50$) in Figure~\ref{fig::informativeness distribution}.
The blue line is the mean value of informativeness, and the shade indicates standard deviation.
The horizontal black line marks the threshold of the soft-selection.
According to the figure, the increasing mean value of informativeness tells us that more recent reading history helps more in expectation. At the same time, the significantly high variance confirms that historical news at each position has the potential to interact with the candidate. Therefore, with a small $K$, Recent($K$) cannot access earlier historical news that tends to be useful in interaction. Rather, SFI is quite able to inspect them and involve them in fine-grained interactions intelligently. Moreover, the informativeness of the history news whose position is farther than $40$ hardly reaches the threshold. So they are considered authentically uninformative and masked even though they are among the top $K\ge 40$. This explains the saturation of SFI's effectiveness and justifies our intuition.

%
\subsubsection{Influence of the Informativeness Threshold}
\begin{figure}[t]
    \includegraphics[width=\linewidth]{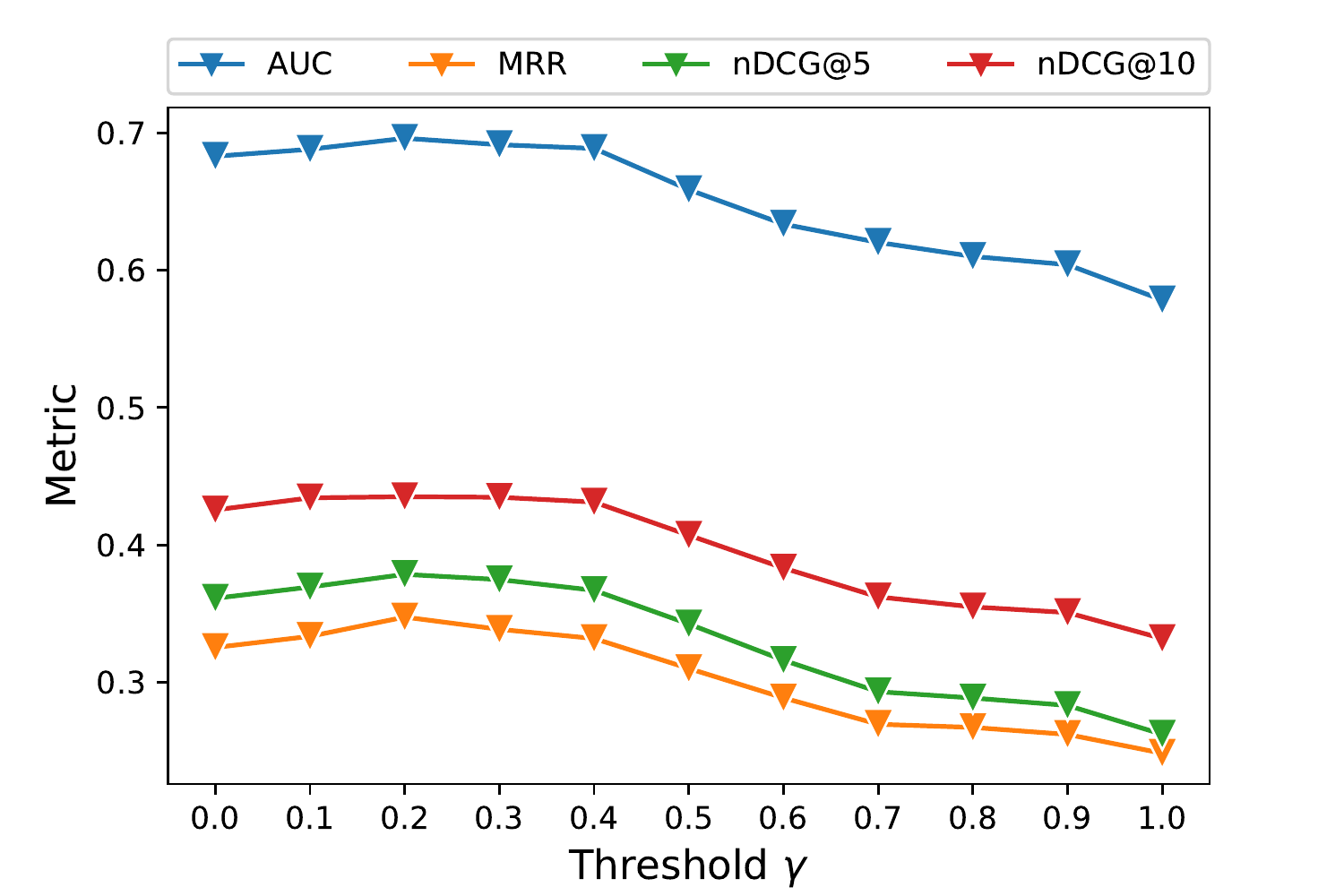}
    \caption{The effectiveness of SFI with different value of $\gamma$.}
    \label{fig::threshold analysis}
\end{figure}
Another crucial factor of SFI is the informativeness threshold in the soft-selection network. The effectiveness of SFI with different threshold settings is shown in Figure~\ref{fig::threshold analysis}. In summary, the threshold shouldn't be too large or too small. When $\gamma< 0.1$, almost all history news articles are considered informative, so the selection fails. When $gamma > 0.3$, the \emph{history selector} rules out too many history news articles, including the valuable ones. The gradient cannot be passed adequately, either. Hence the model's effectiveness declines. When it reaches $1$, all fine-grained representations are masked as $0$, completely disabling the \textit{news interactor}. Recall that the coarse-grained matching signals persist, leading to better results than random recommendation. Overall, $\gamma = 0.2$ is the optimal configuration.

\section{Conclusion and Future Work}
\label{section::conclusion}
Capturing fine-grained interactions brings more accuracy and higher online costs for news recommenders. In this work, we proposed a selective fine-grained interaction framework to select a small number of valuable historical articles for interaction, drawing a good balance between efficiency and effectiveness. With the help of the \textbf{learning-to-select} mechanism, the selection can be performed efficiently, sparsely, and automatically. Experimental results show SFI can significantly improve the recommendation effectiveness by $2.17\%$ over the state-of-the-art models with four times faster speed. We experimented a lot to provide comprehensive insights of SFI and studied the efficiency-effectiveness trade-off it achieves. In the future, we will dig deeper into representing users with terms to further improve the efficiency while keeping the effectiveness.
\clearpage
\bibliographystyle{ACM-Reference-Format}
\bibliography{sample-base}


\end{document}